\documentclass[12pt]{article}

\usepackage[centertags]{amsmath}
\usepackage{amssymb}
\usepackage[pdftex]{graphicx} 
 
\title{Uncertainty In The MAN Data Calibration \& Trend Estimates}

\author{William M. Briggs\\matt@wmbriggs.com\\New York City, NY\\
 \and 
 Jaap Hanekamp \\
 j.hanekamp@ucr.nl; hjaap@xs4all.nl \\
 University College Roosevelt, Middelburg, the Netherlands\\ 
 Environmental Health Sciences\\ University of Massachusetts, Amherst, MA, USA}

\begin{document}

\maketitle

\begin{abstract}
We investigate trend identification in the LML and MAN atmospheric ammonia data. The signals are mixed in the LML data, with just as many positive, negative, and no trends found. The start date for trend identification is crucial, with the trends claimed changing sign and significance depending on the start date.  The MAN data is calibrated to the LML data. This calibration introduces uncertainty never heretofore accounted for in any downstream analysis, such as identifying trends. We introduce a method to do this, and find that the number of trends identified in the MAN data drop by about 50\%. The missing data at MAN stations is also imputed; we show that this imputation again changes the number of trends identified, with more positive and fewer significant trends claimed. The sign and significance of the trends identified in the MAN data change with the introduction of the calibration and then again with the imputation. The conclusion is that great over-certainty exists in current methods of trend identification.

\end{abstract}



\section{Introduction}
\label{intro}

The National Air Quality Monitoring Network or Landelijk Meetnet Lucht-kwaliteit (LML) network  has been measuring ammonia since 1993.  The Measuring Ammonia in Nature (MAN) network is active since 2005 and comprises of a network of atmospheric ammonia monitoring areas with passive samplers at various locations in the Netherlands; see \cite{StoNoo2016,SutDra2015} for descriptions. Both networks play an important role in the Integrated Approach to Nitrogen (PAS). The national plan combines source-measures to reduce nitrogen emission levels and ecological restoration measures. The effectiveness of this plan is monitored with these two network..

There are 6 LML stations that automatically measure ammonia concentrations on an hourly basis; these stations also carry three MAN-like passive samplers. The data from these passive samplers is first averaged and the average is calibrated, i.e. statistically adjusted, so that it better accords with the gold-standard LML data. This adjustment is done on a per-month basis. The method of calibration is given in \cite{LolNoo2015}. We describe the method below, and provide a method to estimate the uncertainty in the resulting calibrated data. This uncertainty has not yet been accounted for in the scientific literature, and is of significant consequence because the calibration results at these 6 stations are applied to all MAN area data. The uncertainty subsequently propagated by this calibration process has yet to be tracked, especially in calculations of trends at the MAN locations. 

Because of the  uncertainty in identifying trends, we re-investigate trends in the LML and MAN data, using all available data. We find that the evidence of decreasing ammonia trends to be as strong as increasing trends, and as strong as no trends. In other words, there are differences in identified trends which depend on the station.

The MAN data is, after the calibration, an estimate of atmospheric ammonia as would be seen by the same sensors used at LML stations. The calibration is thus a prediction of atmospheric ammonia. This prediction has historically been taken as is; i.e., directly from the calibration process. But the calibration is imperfect, and thus so are the predictions, which should not be taken, as they are now, as being certain. So we next estimate the uncertainty added by the calibration process and then carry forward this uncertainty to the trend analysis. We find that certainty in trends identified, as expected, decreases. There is also evidence of positive and negative trends, which are roughly equal in number. Again, there are many differences between stations. 

Finally, we also carry through the uncertainty introduced by the imputation of the MAN data, and discover that it too decreases certainty in trends identified. There are occasional missing data at some MAN areas;  \cite{LolNoo2015} provide a method of imputation for these gaps, which gives a one-time guess of the missing values.  The broad effects of this imputation and the uncertainty introduced by it and the calibration at predicting atmospheric ammonia are described below.

\begin{figure}[htbp]
        \centering
        \includegraphics[scale=.4]{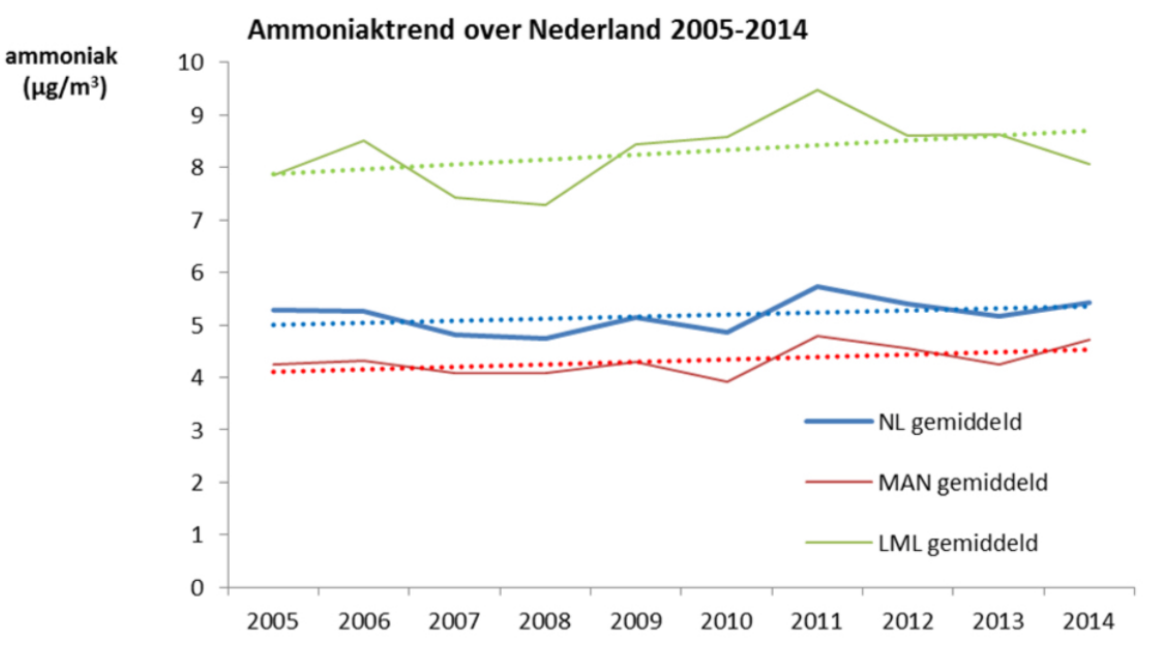}
\caption{\label{fig6SN} Figure 6 from \cite{StoNoo2016}. Ordinary regression lines are over-plotted on yearly atmospheric ammonia levels; see text for details. None of the trends evince traditional statistical significance.}
\end{figure}

It is of great interest to discover whether or not systematic changes are occurring in atmospheric ammonia levels. This is why trend analyses have often been done. For instance, Fig. \ref{fig6SN} reproduced from \cite{StoNoo2016} shows the result of fitting linear regressions to MAN, LML, and overall Netherlands (NL) yearly average ammonia levels.

Identifying what is or is not a trend is of importance. Often, a linear regression is fit to observed data, and the trend coefficient in the model is examined for statistical ``significance", which by tradition is defined as evincing a p-value $<0.05$. None of the estimated trends in Fig. 6 meet the usual definitions of statistical significance: NL gemiddeld $p=0.29$, MAN gemiddeld $p=0.13$, LML gemiddeld $p=0.21$; from Table 2 on page 21 of \cite{StoNoo2016}.  

The authors re-computed Fig. 6 by, they say, normalizing each series by its mean (diving each by its mean). They claim this improved statistical significance (Table 3, p. 22): NL gemiddeld $p=0.05$, MAN gemiddeld $p=0.07$, LML gemiddeld $p=0.07$. This is curious because these results do not accord with regression. Adding to or dividing a set of observations by a constant can of course change the estimate of the trend, but it cannot change its p-value. This causes suspicion that the results in \cite{StoNoo2016} might be in error.  Even if not, dividing a series by a constant cannot create a trend from data where a trend does not exist, however much it might such calculations allow easier comparisons between series of widly different variances.

Another surprise is that the authors of \cite{StoNoo2016} reject the traditional definitions of significance and and claim p-values as high as 0.34 indicate that trends in data are likely; see their table on page 9 of \cite{StoNoo2016}. This is expanded in \cite{KruHoo2018} (p. 32) up to p-values of 0.66, which are said to indicate ``Trend ongeveer net zo waarschijnlijk als niet" (``Trend equally likely as unlikely").   

These definitions of significance are particularly odd in the face of the latest statistical research in which prominent statisticians and scientists are calling to {\it retire} statistical significance as a concept, e.g. \cite{AmrGre2019}. The American Statistical Association issued a statement warning about excesses due to reliance on p-values; \cite{Was2016}. Others call for a {\it lowering} of the usual threshold from 0.05 to 0.005, e.g. \cite{BenBer2018,BerSel1987}. These authors say that (emphasis original) ``Associating `statistically significant' findings with P $<$ 0.05 results in a high rate of false positives {\it even in the absence of other experimental, procedural and reporting problems}."  No statistical authority anywhere endorses the view of calling a hypothesis ``equally likely as unlikely" for p-values up to 0.66, or even anywhere larger than 0.1. 

It is also a common technical mistake to use the p-value to infer a trend is ``likely" or otherwise present, as \cite{Giger2004} shows. P-values are in no way related to the probabilities hypotheses, like trends present or absent, are true or false. \cite{GibGib2004} specially criticize the practice of p-values to describe ``almost significant" research, i.e. claiming an effect when p-values are larger than the traditional measures of significance. \cite{ArmKor2017} open their abstract stating that according to the American Statistical Association the ``widespread use of `statistical significance' as a license for making a claim of a scientific finding leads to considerable distortion of the scientific process." In similar analyses of trends on MAN data, \cite{VanWich2017} (p. 357) say ``The normalized trend for all monitoring stations of 19\% is almost significant on a 95\% CI as the p value of 0.07 shows," acknowledging the accepted value of 0.05.

Many other authorities call for a complete abandonment of p-values as evidence, e.g. \cite{McSGal2017,TraAmr2018,Bri2019a,HebLin2008,Bri2016,ZilMcc2008}. These authors show how easy it is to generate false positives with p-values even less than 0.05. These unwanted consequences necessarily become much more likely with a threshold of 0.34 or higher. 

Because of this, and because use of p-values is still common, we shall adopt the usual threshold of 0.05 when discussing trend identification.

We begin first with the calibration of the MAN data, and provide methods to assess the uncertainty introduced in the process. We next do a trend analysis of the LML data. Since the LML data goes back farther than the MAN data, we can examine how varying the start date of a trend analysis affects results.  Finally, the trend analysis in the MAN data, in its raw, calibrated, and calibrated and imputed forms are carried out and the results compared, with the full uncertainty of the methods incorporated into the trend estimates.

\section{Calibration}
\label{scali}

For each month, data is collected at each of the 6 LML stations. Between 1993 and 2015, AMOR was used to measure hourly ammonia concentrations; since 2016 the so-called miniDOAS (Differential Optical Absorption Spectroscopy) is used. Besides the normal LML samplers, three additional passive samplers are attached at each LML station. These represent the same kind of samplers used at MAN locations. The mean of the three values provides the estimate of MAN-like data at the individual LML stations. At each month, this MAN-like data is next calibrated, i.e. statistically adjusted, so that it better resembles the LML data. The LML data is taken as the gold standard, as it were.  The calibration results from these 6 stations is then applied to all MAN locations at the same month so that data at those locations better represents actual atmospheric ammonia. In this sense, the calibration is a prediction of what the LML measurement would have been had the LML measurement taken place at these stations.

The calibration of the MAN-like data at the LML stations proceeds as follows, as given in \cite{LolNoo2015}.  For each month, the 6 three-sample mean MAN values ($x$) are regressed against the LML divided by MAN values ($y$). I.e. 
\begin{equation}
\label{cal1}
y = a + bx + \epsilon,
\end{equation}
where $a$ and $b$ are the (monthly) calibration coefficients, and $\epsilon$ expresses uncertainty in the measurement differences of the passive and LML samplers. Ordinary linear regression is used to estimate $a$ and $b$. Ideally, if the passive sampler means exactly matched the LML samplers, $y$ would everywhere equal 1; thus the ideal model coefficients, for perfectly calibrated data, would be $a=1, b=0$. Departures from these values indicate increasingly uncalibrated MAN passive sampler values.  To correct the MAN values, the calibration equation (\ref{cal1} ) is inverted using the estimated coefficients to produce (recalling $y = \mbox{LML}/x$)
\begin{equation}
\label{cal2}
x_c = (\hat{a} + \hat{b}x)x,
\end{equation}
where $x_c$ represents the calibrated values. As is easily seen, if $\hat{b}=0$ and $\hat{a}=1$, as in the ideal situation, $x_c = x$, i.e. unmodified values. 

\begin{figure}[htbp]
        \centering
        \includegraphics[scale=.5]{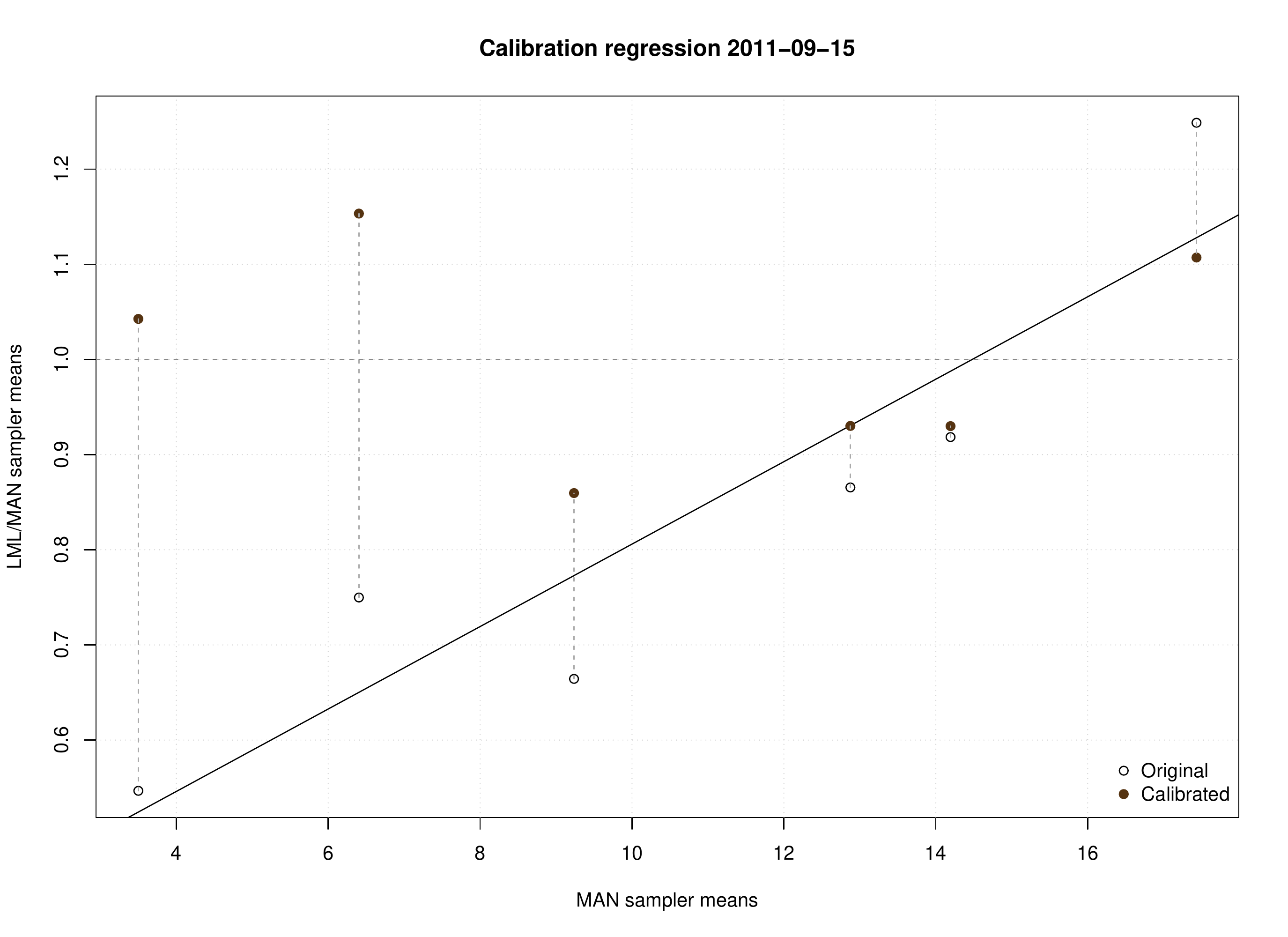}
\caption{\label{fig1} The calibration exercise for 15 September 2011. The open circles are the MAN samplers means. The y-axis is $Y$, the LML divided by the MAN samplers means. The regression line is solid black. The calibrated MAN values are computed and then plotted as closed dark circles. A dashed vertical line joins them to show how much of an effect the calibration had. The light horizontal line at $y=1$ shows the ideal, i.e. perfectly calibrated values.}
\end{figure}

Figure \ref{fig1} shows the calibration exercise for 15 September 2011. The open circles are the MAN samplers means, indexed on the x-axis. The y-axis is $y$ from (\ref{cal1}), the LML divided by the MAN samplers means. The regression line is solid black. The calibrated MAN values are computed via (\ref{cal2}) and then plotted as closed dark circles. A dashed vertical line joins them to show how much of an effect the calibration had. The light horizontal line at $y=1$ shows the ideal, i.e. perfectly calibrated values. If the calibration were perfect, all the dark circles would lie on the $y=1$ line.

\begin{figure}[htbp]
        \centering
        \includegraphics[scale=.5]{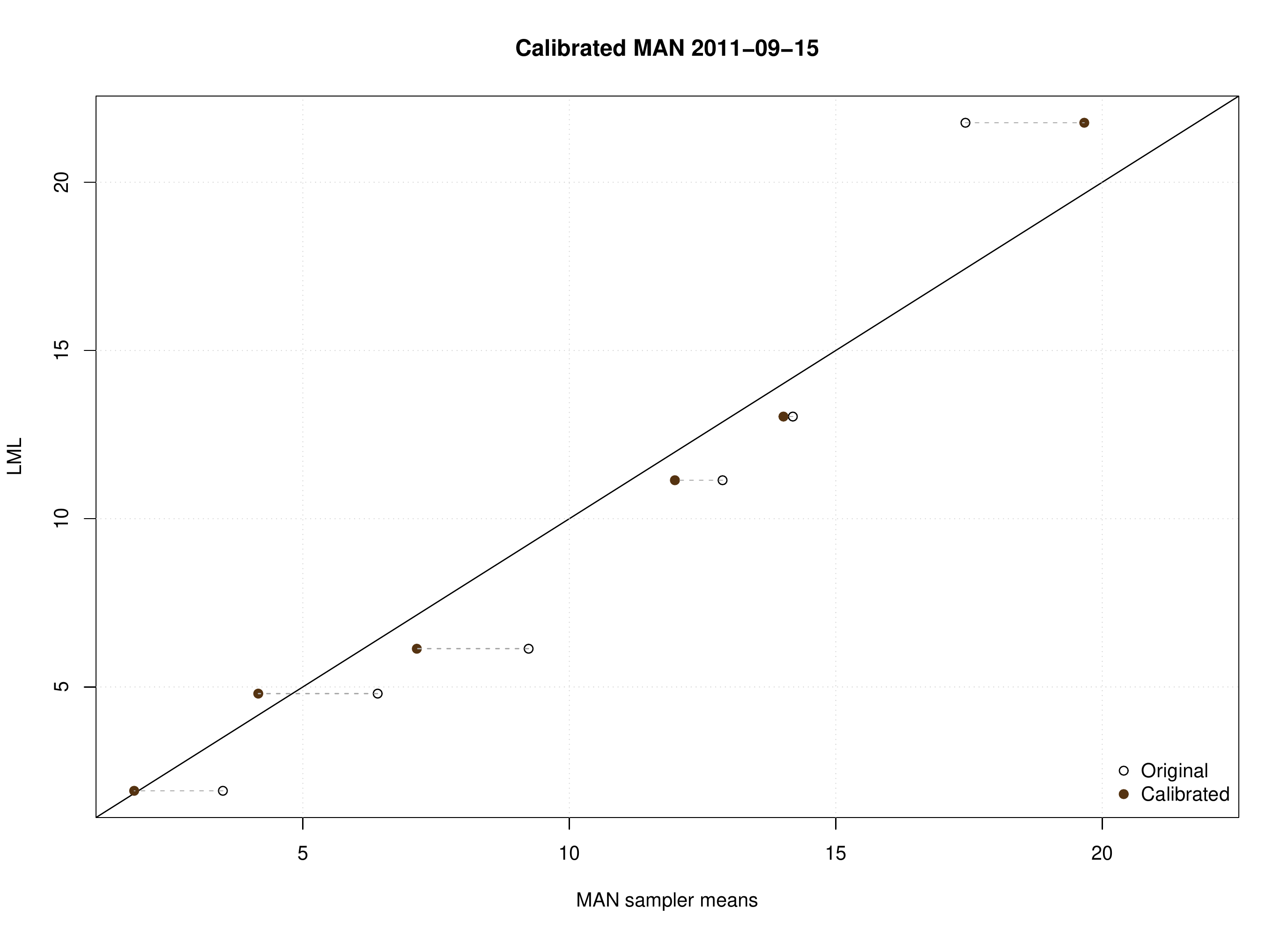}
\caption{\label{fig2} The original and calibrated MAN data from Fig. \ref{fig1}. The open circles are the original MAN samplers means. The dark circles are the calibrated values. The y-axis are the LML values.  A dashed horizontal line joins them to show how much of an effect the calibration had. A one-to-one line is over-plotted.}
\end{figure}

Figure \ref{fig2} shows the original and calibrated MAN data from Fig. \ref{fig1} in a native view. The open circles are the original MAN samplers means. The dark circles are the calibrated values. The y-axis are the LML values.  A dashed horizontal line joins the pre- and post-calibration values to show how much of an effect the calibration had. A one-to-one line is over-plotted. If the calibration were perfect, each of the dark circles would lie on the one-to-one line. It is not clear from this example, where every point has moved closer to the line after calibration, but calibration can cause some values to become worse, i.e. move away from the line.

The calibration process is predictive. The MAN passive samplers attempt to predict, in a physical-chemical way, the value of the more sensitive LML samplers, which are taken as the ``gold standard" of atmospheric NH3. The statistical calibration, on average, moves the MAN sampler data closer to the LML data. Ideally, the calibration procedure would allow as inputs MAN passive sampler data, and produce an output equally exactly the LML values. This ideal situation is obviously not plausible for a regression with only 6 points (and historically sometimes only 5). Indeed, Fig. \ref{fig2} proves this because not all calibrated points line on the one-to-one line.

\begin{figure}[htbp]
        \centering
        \includegraphics[scale=.5]{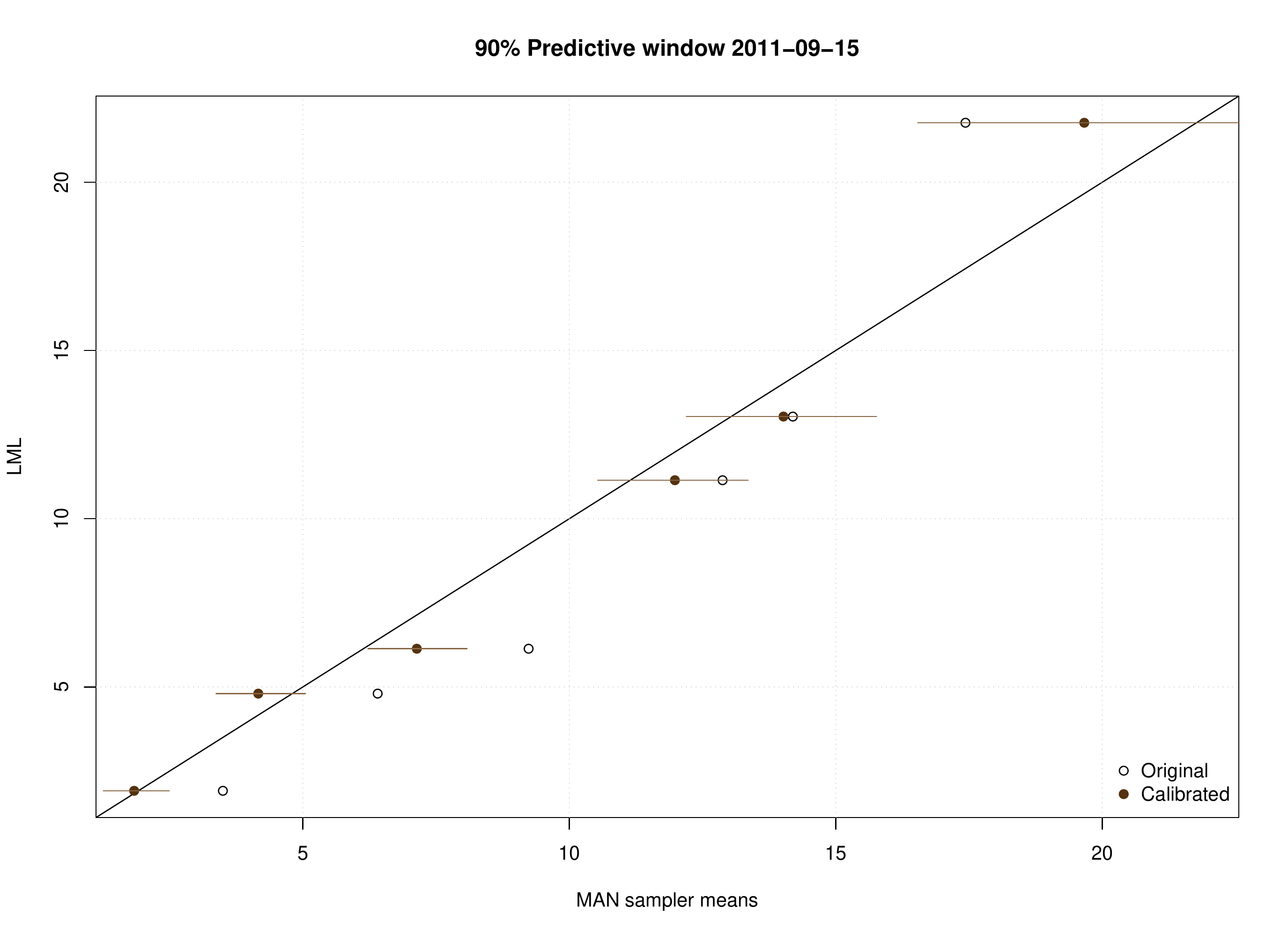}
\caption{\label{fig3} The 90\% calibration predictive intervals for MAN data from Fig. \ref{fig1}. The open circles are the original MAN samplers means. The dark circles are the calibrated values. The y-axis are the LML values.  Dark horizontal lines indicate the 90\% intervals for $x_c$, accounting for the uncertainty in $a$ and $b$ in (\ref{cal1}). A one-to-one line is over-plotted.}
\end{figure}

The imperfection means there is some uncertainty in the predictions of atmospheric ammonia, i.e. in the calibrated values as predictions. Fig. \ref{fig3} shows this uncertainty for the same data in Fig. \ref{fig2}. The regression (\ref{cal1}) was redone, this time using the \verb|stan_glm| method from the R package \verb|rstanarm|, using default priors. The estimated values of $a$ and $b$ were then used to compute predictive versions of $x_c$, via the predictive posterior distributions. The 90\% interval of each calibrated MAN value was plotted as a dark horizontal line; the mean estimated $x_c$ is again a dark  circle, which is the same as before. The uncalibrated values are open circles also as before. This nicely accounts for the uncertainty inherent in the calibration. The interpretation of each horizontal line is that, given the data and calibration model, there is a 90\% chance the true value of atmospheric ammonia (as measured by LML) will lie in this interval. As expected, most of these intervals overlap the one-to-one line.  

These are not confidence or credible intervals, but predictive intervals. The confidence or credible intervals would be calculated for the coefficientes $a$ and $b$, to express uncertainty in their value. However, we are interested not in the model coefficients {\it per se}, but in actual ammonia values. Thus we want the prediction interval of the calibration process: we want the uncertainty in the guess the calibrated MAN data is making of actual atmospheric ammonia. See, for example, \cite{Gei1993,JohGei1982,Har2001} on prediction intervals in regression.  

Again, these intervals say there is a 90\% chance the true value of NH3 lies in the stated interval, given the model and MAN-like measurements.  Most of the intervals overlap the one-to-one line, so with some caveats the calibration model does a reasonable job. It is difficult to see in this plot, though we demonstrate it more clearly next, that the predictive error increases with size of the original MAN value. What is clear here is that the uncertainty, while reasonable, can be substantial, a matter of several $\mu$g m$^3$.  This is important when trends with sizes around this average are claimed. Reiterating, the claims made with the MAN data have far-reaching policy consequences.

\begin{figure}[htbp]
        \centering
        \includegraphics[scale=.6]{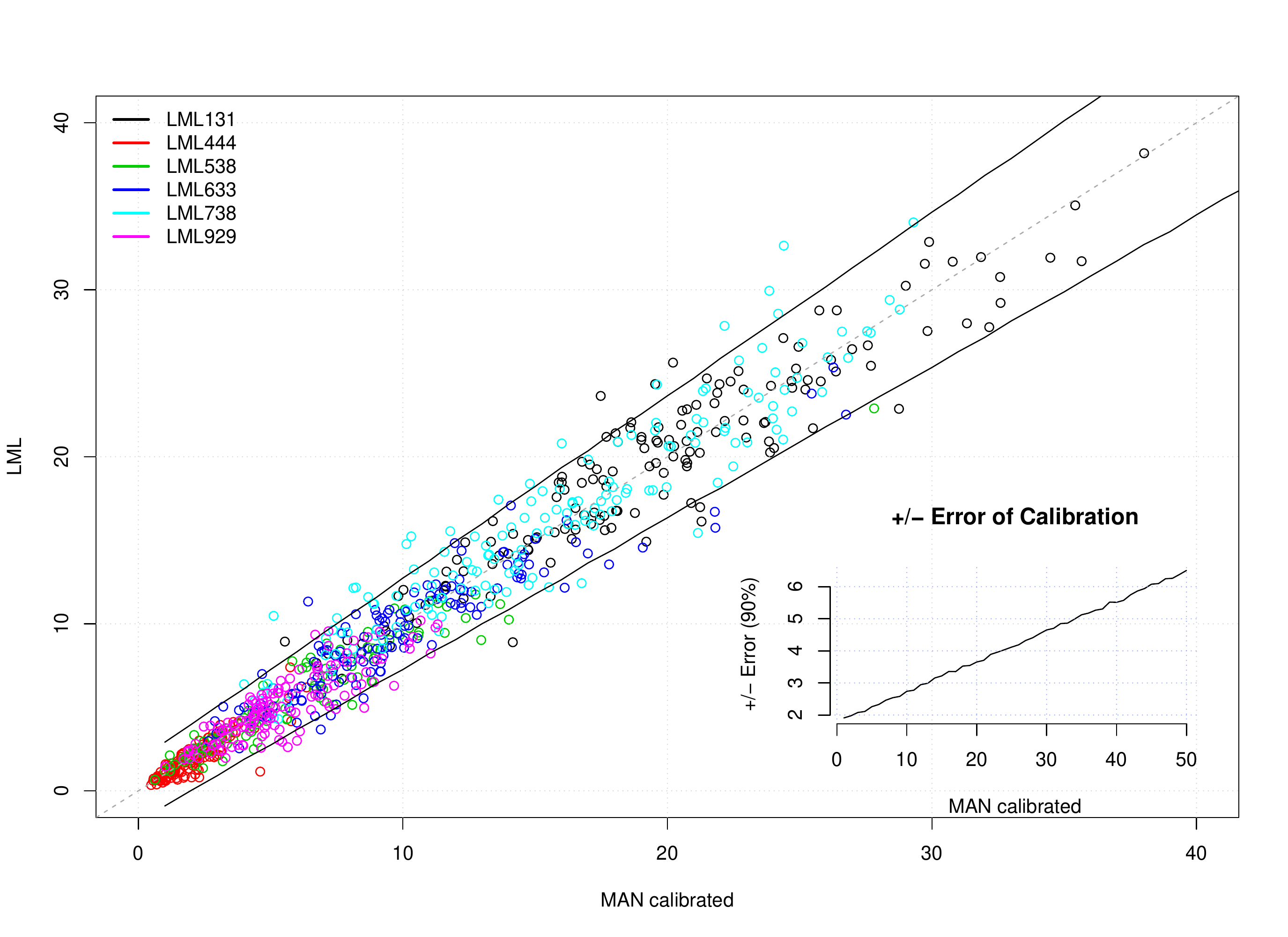}
\caption{\label{fig4} This shows the calibrated MAN data against the original LML data, across all stations and months. A light grey dashed one-to-one is given for reference. The two solid black lines are the approximate 90\% predictive intervals. In other words, given a particular MAN calibrated value, these line encompass the 90\% chance the true LML value will fall in the interval. An inset plot shows the mean error introduced, or left over, in calibration, of MAN predicting LML data. It is clear the size of the error depends on the level of MAN data, with higher levels providing greater uncertainty.}
\end{figure}

It is of interest to estimate the average predictive uncertainty, or predictive error of the calibration.  Shown in Fig. \ref{fig4} are all 155 months with each of the 6 (or 5) LML station values plotted against the calibrated MAN values, using (\ref{cal2}). A dashed one-to-one line has been over-plotted.   The two solid lines are approximate 90\% prediction bounds for a regression model to be explained in a moment. It can been see the calibration does a good job in the sense that there is no consistent negative or positive bias. Even so, uncertainty does increase with MAN values.

Again, any given calibrated MAN value is not likely to predict the LML (hence atmospheric NH3) value perfectly, but will have some error. This error, as the figure shows, increases with the value of the data at the MAN-like samplers.  Because there is no systematic bias, the error is practically the same in either direction, too high or too small. This suggests an error regression model is apt, where the absolute value of the difference of LML and calibrated MAN values was regressed on the calibrated MAN values. Because there is no inherent bias, the slope is set to 0. In other words, we estimated (\ref{cal1}) again, this time using \verb|stan_glm|.
    
\begin{equation}
\label{cal3}
|\mbox{LML} - x_c| =  cx_c + \epsilon.
\end{equation}

Solely for illustration, a plot using this regression to estimate the 90\% prediction error bounds by $x_c$ is given in the inset to Fig. \ref{fig4}.  For instance, whenever we see the MAN calibrated value of 30 $\mu$g m$^3$, the true LML value is 90\% likely to be in the interval $30 \pm  5$ $\mu$g m$^3$. With MAN calibrated values of 10 $\mu$g m$^3$, the true LML value is 90\% like to be (about) $10 \pm  3$ $\mu$g m$^3$.  There is not much difference between this method and the more usual errors prediction, i.e. using the variance of the predictions from the regression (\ref{cal1}), which we indeed use below. But (\ref{cal3}) allows a refinement of the variance based on MAN level.  The interval at low values of MAN is too wide, in the sense that more than 90\% of the actual values are in the window (meaning for the smallest MAN values, this esitimated prediction interal is closer to, say, 98\% than 90\%). This means that the linearity in increasing error assumption is somewhat flawed. However, since this graph is only for illustration purposes only, this is not crucial. 

It is clear from the inset figure that the prediction error is not insubstantial, representing at least 10\% of the size of MAN calibrated values.  This kind of uncertainty should and must be incorporated into any downstream analysis which seeks to account for actual atmospheric ammonia.  Since all MAN station data (and not just these 6 used for calibration) are calibrated, and MAN data are used to assess atmospheric ammonia and trends in the same, the uncertainty in the values should be accounted for.  We suggest methods how this might work in the next section on trend analysis.

\section{Trend Setters}
\label{trend}

Asking whether there has been a trend in a time series is not a simple question, though it is often taken to be. We must first decide what a ``trend" is.  See \cite{Bri2016}, Chapter 10, for a complete discussion. 

Trend definition is the first difficulty. A second and serious problem is due to so-called edge effects. Choosing different start and end points of a time series can result in different decisions about whether a trend is present. This happens frequently using monthly data, when the signal possess any kind of regularity, and where the regularity, such as seasonality, is not accounted for. For a crude example, looking for a trend in monthly temperature data that started in January could lead to a different result if the start date were July, instead. Two different conclusions can be reached just by clever choice of the start and end points.

Strict linear signals are the most common definition of a trend. If this linear signal is real, it is thought some constant linear force is causing data values to rise or fall, a force that is separate from other causes of changes in the data. When a linear trend has been identified through statistical algorithms, this constant cause or strict linear force is thought to have been verified, even though statistical models cannot identify cause (``correlation is not causation"). This is all very odd because, except in rare instances, this is not how causes work in physical series.  It cannot be emphasized too strongly that the atmosphere, or anything in contact with it, does not experience the trend: the atmosphere experiences the actual data values.  Trend identification, therefore, is far more important for forecasting purposes than for explaining what happened. 

The most common definition of trend is a positive slope identified via an ordinary linear regression of time on observation. This trend is often unphysical, i.e. only a line drawn over some data and not representative of an underlying linear cause, and its misuse leads to mistakes in causal ascription. Nevertheless, because it is common, we will use it to demonstrate trend estimation techniques. 

\subsection{LML}

\begin{figure}[htbp]
        \centering
        \includegraphics[scale=.6]{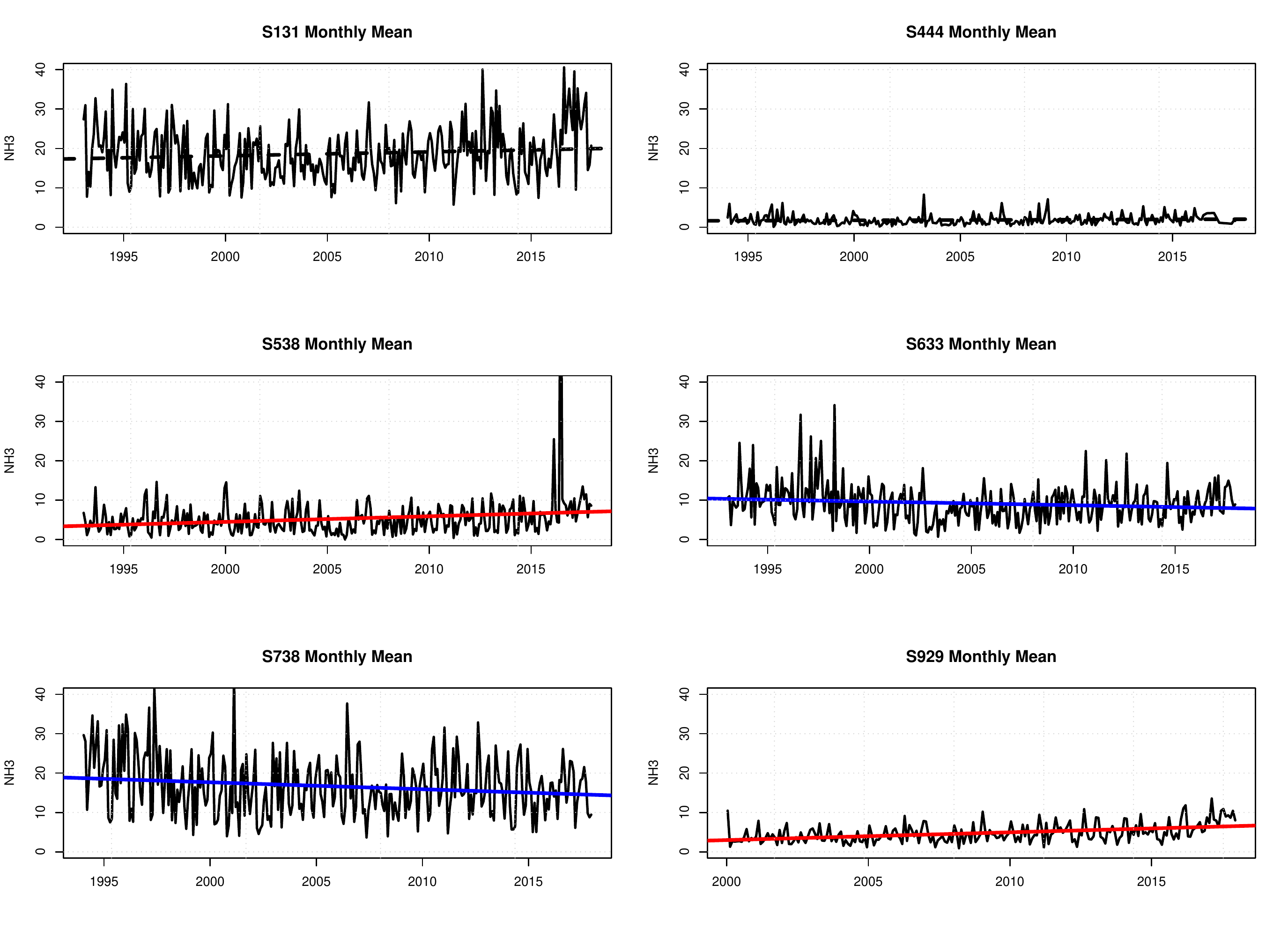}
\caption{\label{lml1} The monthly data of each LML station, with a regression trend line over-plotted. The line is dashed black for no significant trend, red for significant increasing trend, and blue for significant increasing trend. The scale is identical for each station, for ease of comparison. The spike at station S538 is explained in the text.}
\end{figure}

For simplicity, and because the MAN data are calibrated to the LML, we begin with the 6 stations of the LML data. The stations are: Vredepeel (S131), De Zilk (S444), Wieringerwerf (S538), Zegveld (S633), Wekerom (S738) and Valthermond (S929). Figure \ref{lml1} shows the monthly means of NH3 from January 1993 through December 2017 for each station where data is available.  Ordinary linear regressions by time were calculated and over-plotted. The lines are dashed if the regression did not evince a wee p-value (less than the traditional 0.05); they are blue for negative trends and red for positive trends with wee p-values.  Two stations, 131 and 444, show no trend (by the given definition), both 633 and 738 show negative trends, and 538 and 929 showed positive trends. 

The June 2016 value of NH3 at station 538 reached a high of 82.8 $\mu$g m$^3$. This appears genuine. It originates in the hourly data, which saw measures of 137  $\mu$g m$^3$ starting 2016-07-14, 15:00, rising steadily to a peak of 1198 $\mu$g m$^3$ on 2016-07-15, 06:00, falling back steadily throughout that next day, and again the next two nights peaking and falling in much the same way.  As can be seen in the February spike also in 2016, this is not the first time it happened. Since these were the felt or experiences values of NH3, they are not ``outliers" and so they are used in the analyses below.

All students of regression will understand that these very large monthly values at this station, roughly 8 to 10 times larger than the remaining observations, carry great weight in the results. Yet since the data is (or appears to be) genuine, it cannot be removed: again, these were, after all, the values of NH3 experienced. But re-running the regression stopping at January 2016 shows the trend ``disappears", i.e. the p-values associated with the regression is no longer less than 0.05. This is our first example that trends come and go depending on start or end points.

The ``trend" is there or is not depending on this statistical definition. But it is clear that the data themselves do not cooperate in this assessment over their ranges. The data values, of course, sometimes rise, sometimes fall: the variability is not small.  What is causing these undulations is not a question that can be answered via statistical modeling. The best we can do with statistical models is to make predictions of future values. We do not need the models to tell us what did happen, for we can plainly see what did or did not happen. The regression lines did not happen. The data did. 

Now if there is an unknown linear force that causes a trend that is identified by statistical means, then for this data there there necessarily had to be four such distinct causes. Each operated differently at different areas across the Netherlands. Two causes operated to increase atmospheric ammonia, and two causes operated to decrease atmospheric ammonia.  Plus, two other areas had no linear causes at work. There has to be a physical or chemical reason for these differences if the trends are real.  While it is not impossible these forces exist, or are absent at some locales, it is a strain against common sense to believe it.

There is no justification for combining these 6 stations to get some sort of overall picture. The observations are too disparate, for one. Whatever is happening physically at S929 is radically different than what is happening at S738 (the two stations at the bottom row).  Any combination would be averaging already acknowledged different causes. If it is believed there were these different strict linear causes operating in different areas, then aggregating the data is not physically justified.  NH3 values vary greatly from region to region, in magnitude and variability.  

\begin{figure}[htbp]
        \centering
        \includegraphics[scale=.6]{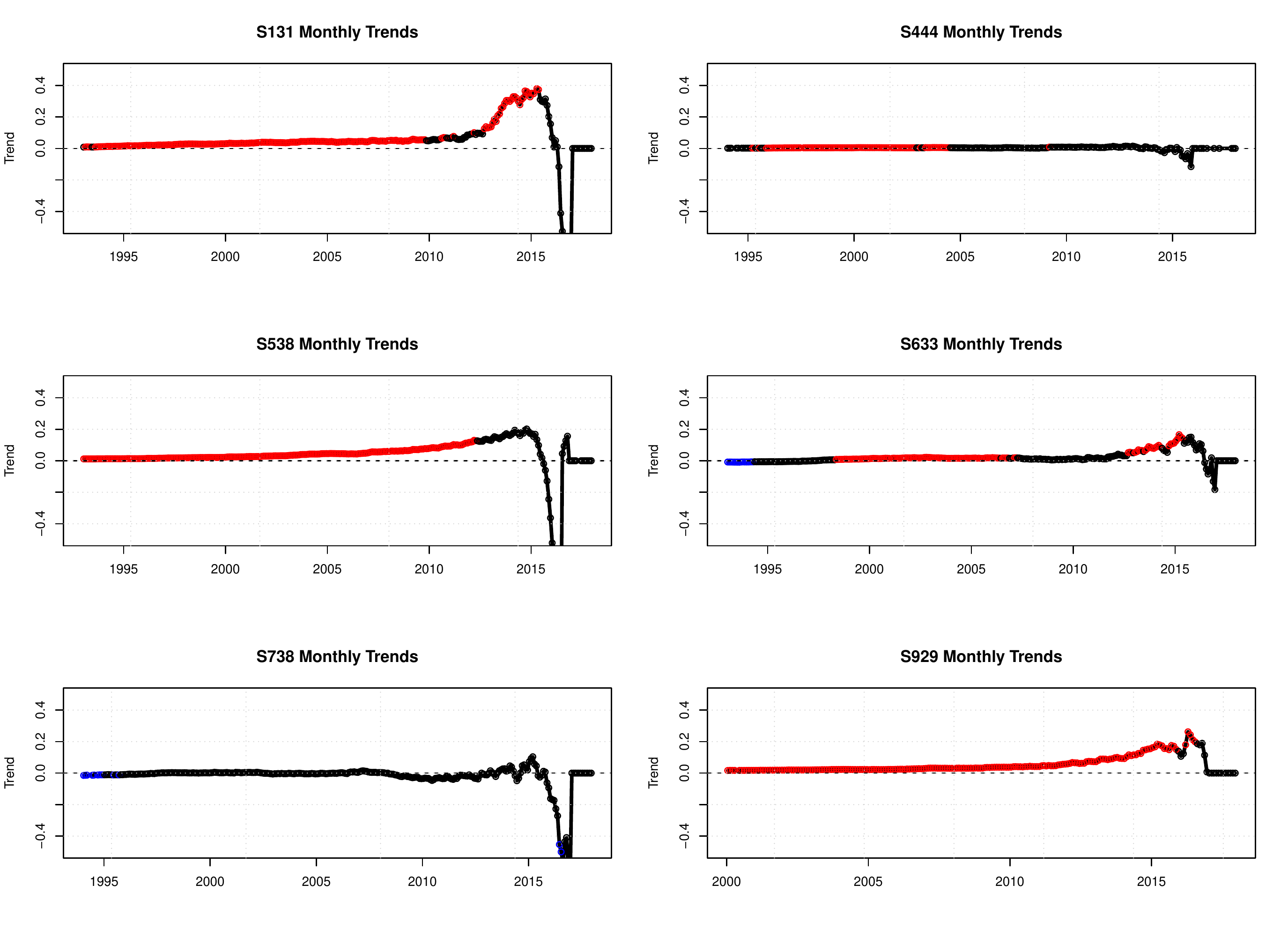}
\caption{\label{lml.sd} This graph shows how varying the start data of the trend analysis affects the resulting trend identified. Each successive month, from month 1 to all but the last year available, was excluded and the trend regression redone. The value of the estimated monthly trend coefficient is shown, in black for non-significance, red for significant and positive and blue for significant and negative. It's clear some stations have both ``significantly" increasing {\it and} decreasing trends, proving both the start date matters and that trends are more complicated than just straight lines imposed on the data.}
\end{figure}

We next demonstrate starting date matters in analyzing trend. At each station, for every month $i$ in Fig. \ref{lml.sd} starting with month 1, we do a regression from month $i$ to the last time point available, and then plot the estimated trend coefficient: red is positive and significant, blue is negative and significant, otherwise black. This procedure is repeated for start times $i = 2, 3, \cdots$, up until the last month - 12. We stop at the last year's worth of data on the assumption there is no point in doing a regression over just one year, i.e. 12 months, of data.

This plot shows the start date matters: the trend comes and goes, and is sometimes negative, depending on the station. Most of the estimated so-called trends are small, at best. If we accept the statistical identification of trend discovers linear cause, then we have to explain how the linear cause changes so rapidly in magnitude and sign, and at certain dates (why these dates and not others, for example). Or we could use this graph to illustrate the difficulty of understanding what trends are.

\begin{figure}[htbp]
        \centering
        \includegraphics[scale=.6]{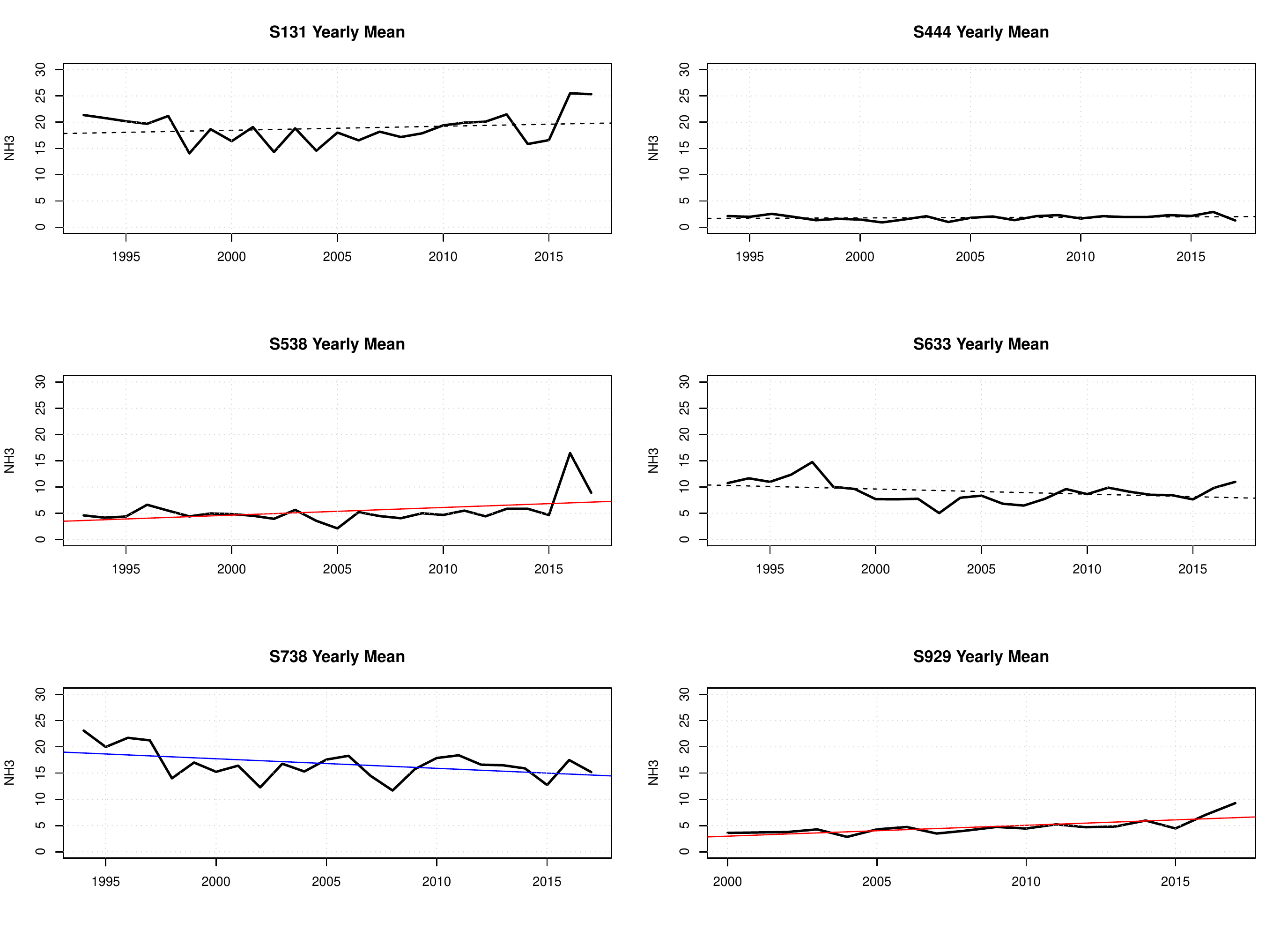}
\caption{\label{lmly} Yearly mean values of NH3 with trends computed as in Fig. \ref{lml1}.}
\end{figure}

Lastly, we repeat the main trend analysis for yearly means, as illustrated in Fig. \ref{lmly} 

Interestingly, the only real change from the montly analysus is the negative trend identified in station 633 in the monthly data, which has ``disappeared" in the yearly data.  Of course, it is well known that p-values are somewhat dependent on sample size, and the yearly means are 1/12th the size of the monthly means.   The spike in station 538 was discovered above to be present in the hourly values, but a naive analysis looking just at yearly values might misascribe the bump to some overall yearly cause. This is another strong reminder that understanding caution is different than statistically claiming trends.

\subsection{MAN}

We now move form the 6 LML station to the MAN data, which consists of 294 measurement locations with at least 5 observations. We use data from 2005 until January, 2018.  The MAN data consists of raw data, data which has been calibrated as above, and data which has been calibrated and where the missing values (if any) are supplied by imputation. The imputation method is described in \cite{LolNoo2015}. We do not attempt to reproduce it here, but use the data to which the imputation has already been implied.  In other words, we take each of these three data sets as they have been given to us, and ask questions about the presence of trends (indicated by statistical methods) to demonstrate that the calibration and imputation processes make large differences in trend identification. 

For each of the three data sets---raw, calibrated, calibrated plus imputed---we computed the trend at each of the 294 stations, noting whether it was positive or negative, and whether the p-values were ``significant" or not. This was done to see how each successive data manipulation influenced the trend estimates. 

\begin{figure}[htbp]
        \centering
        \includegraphics[page=1,scale=.5]{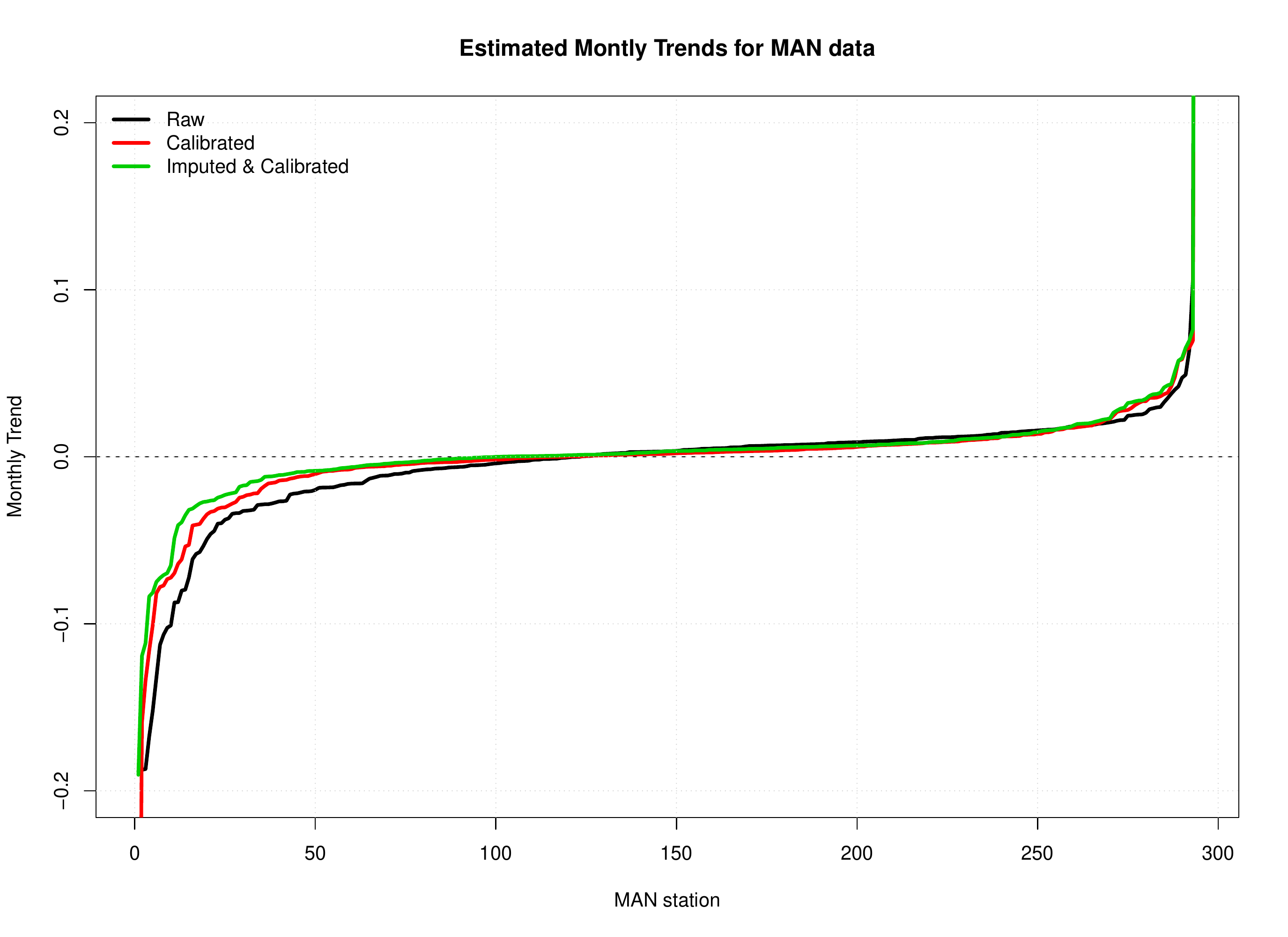}
\caption{\label{fig11} Monthly trend coefficients were estimated from the raw, calibrated, and imputed and calibrated MAN data. In each case the coefficients were sorted from low to high, and the results plotted. The imputed and calibrated MAN data had the smallest negative trend estimates, shown by the green line being above the red and black for negative values (recalling this is not a station-by-station comparison). Yet the imputed and calibrated MAN also had quite a few more larger positive trend estimates: the green line is slightly higher than the red and black lines for middle range values, though all three stations tended to have the same highest estimates. }
\end{figure}

For the first analysis, we calculated and then sorted the trend coefficient for each of the three data sets and plotted them in Fig. \ref{fig11}. The station numbers are arbitrary.  Each point on the lines cannot be compared directly: comparisons will be done below. This plot is only to examine the distribution of trends for each data set.

The trends in the raw data are represented by black lines, calibrated data by red lines, and calibrated and imputed by green.  The raw data had stronger negative trends than the calibrated or calibrated and imputed. The positive trends are similar, with some slight increases in calibrated and calibrated and imputed. This suggests the calibration and then imputation tend to produce on average smaller negative trends, while also producing somewhat larger positive trends. This supposition is examined in detail below, and in the next graph.

\begin{figure}[htbp]
        \centering
        \includegraphics[page=2,scale=.5]{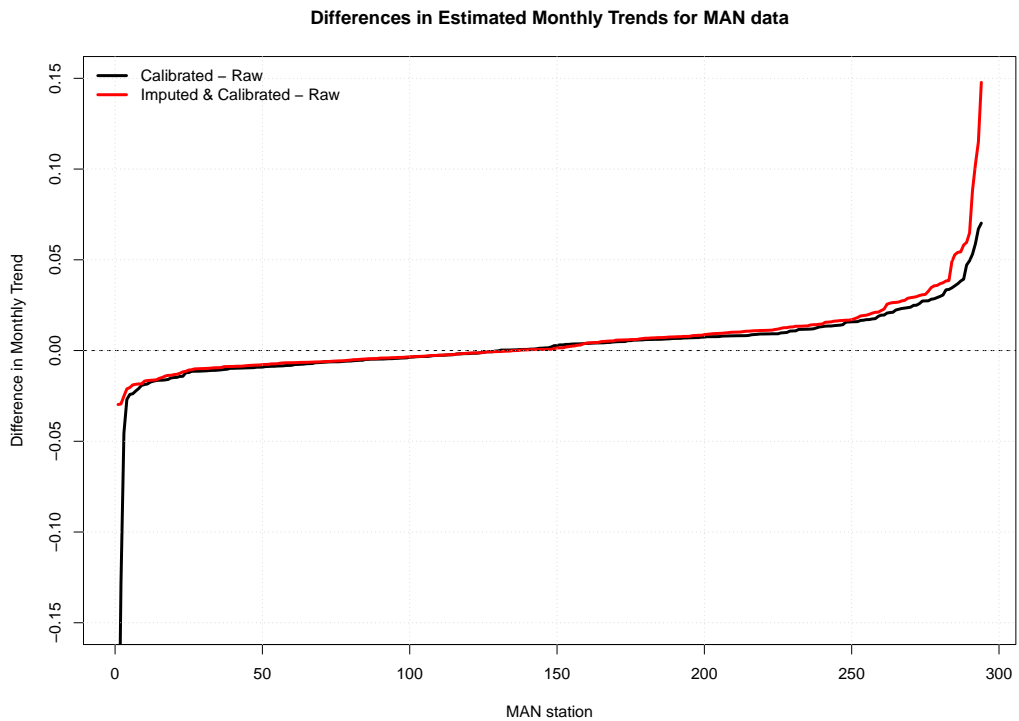}
\caption{\label{figt2} This is similar to and clarifies Fig. \ref{fig11}, except that this is a station-by-station difference of monthly trend coefficients. It shows the raw data had larger negative and smaller positive trend coefficient estimates than either the calibrated or imputed and calibrated estimates. The calibration and imputation are introducing somewhat larger (in absolute value) coefficients.}
\end{figure}

Differences in the trend coefficients at each station were next calculated, as shown in Fig. \ref{figt2}. This graph clarifies \ref{fig11}. Calibrated minus raw is in black, and imputed and calibrated minus raw is in red. Direct comparisons between data sets can now be made. The data are sorted from low to high raw trend coefficients, so station number is again arbitrary. Negative differences indicate stations where the trend in the raw data is larger than in the modified, and vice versa. Differences where the modified data showed larger trends were more evident, as seen in the right portion of the figure. 

Imputation is creating slightly smaller negative trend estimates, but larger positive trend estimates than calibration alone. The differences are many and complicated, proving, as we shall have occasion to emphasize, the calibration and imputation are important, and can cause decisions to be changed. We further illustrate this with a series of tables comparing the trends.

\begin{figure}[htbp]
        \centering
        \includegraphics[scale=.35]{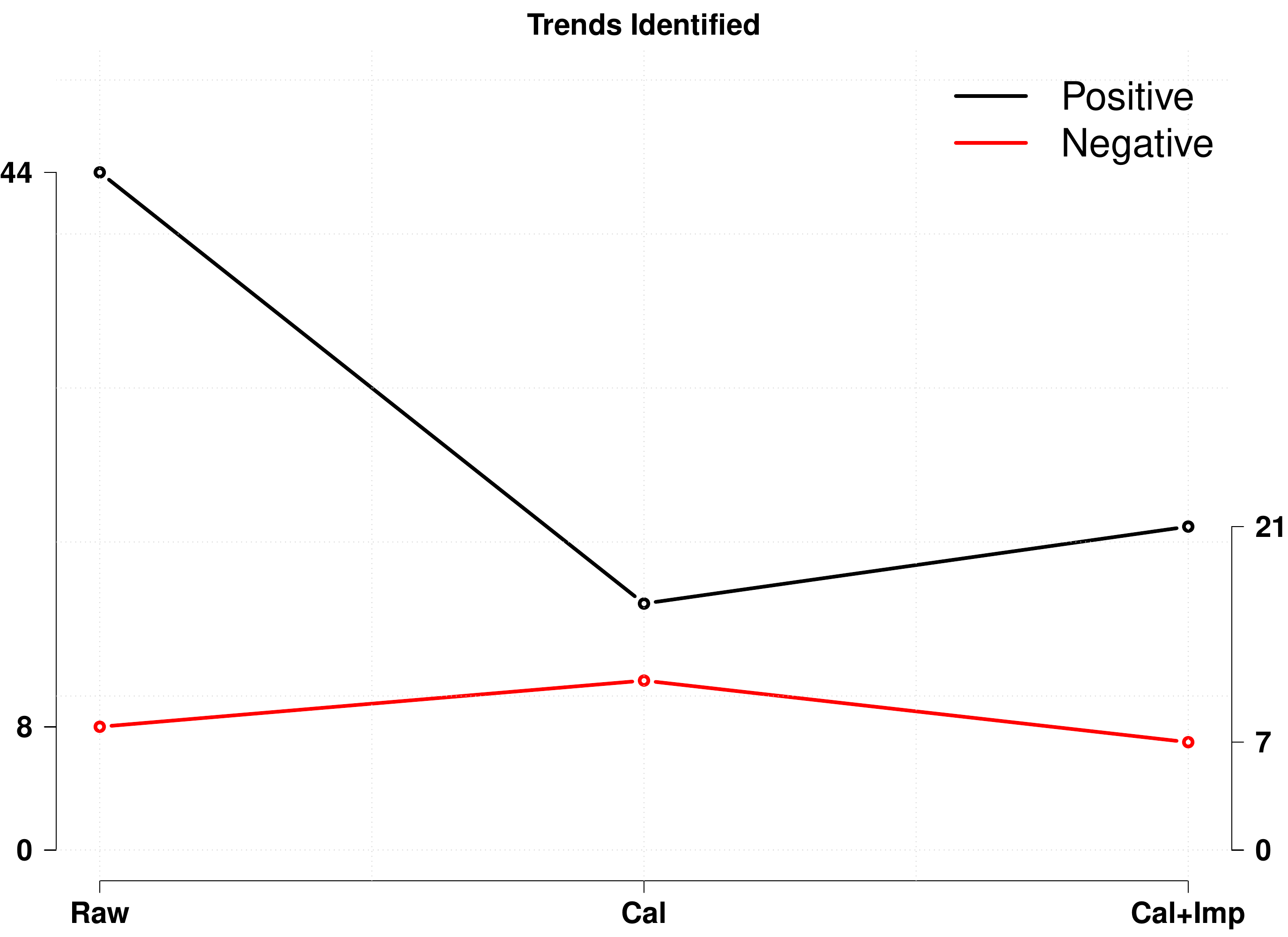}
\caption{\label{figtab1}  The number of positive (black) and negative (red) significant trends identified in the raw, calibrated, and calibrated and imputed data using $p<0.05$.  The imputation causes more postive and fewer negative significant trends. This analysis does not account for the uncertainty introduced in the calibration process, which decreases the number of identified trends. For that, see below.}
\end{figure}

Table \ref{tab1} and Fig. \ref{figtab1} shows the number of positive and negative trends identified in each of the three data sets. The asterisk indicates trends where the p-value was less than 0.05. The raw data had the fewest positive trends, whereas the calibrated and imputed had the most, a 13\% increase.  Negative trends were more frequent in the raw data, with the calibrated and imputed showing the least, a 23\% decrease. 

Raw data did have the most ``significant" trends, which were cut back a great deal in the modified data. This is not surprising when we recall the calibration method ``squashes" the raw data, i.e. reduces some of its variability. Since wee p-values are driven by larger values, it is clear the raw data simply had more of them. We remind the reader raw values are not used in practice, so this analysis is only illustrative of the effects of calibration and imputation. We only want to demostrate there wre differences between the calibrated and calibrated and imputed data sets.  In particular, the number of positive ``significant" trends increased by about 25\%, which shows the imputation process is important. It is not clear whether this is because the imputation process is supplying large values on average, or just more of them. A known weakness of p-values is that as the sample size increases, p-values shrink, even in the absence of true effects. 

{\small
\begin{table}[htbp]
  \begin{tabular}{lccc}
Trend       & Raw & Cal. & Cal. + Imp. \\\hline\hline
+  & 171 & 175 & 194 \\
--  & 123 & 119 & 100 \\
$+^*$  & 44 & 16 & 21 \\
$-^*$  & 8 & 11 & 7 \\
  \end{tabular}
\caption{The number of positive (+) and negative (-) monthly trend coefficients for the three data sets, and also the number of positive or negative significant (*) trends, after accounting for the manipulations introduced by the calibration and imputation processes. The calibration and imputation methods are responsible for fewer trends, positive or negative, being identified in comparison to the raw data.}
\label{tab1}
\end{table}
}

In Table \ref{tab2} we look at these differences more closely. For each row, there are two groups of data, the sign of the trend, and its statistical ``significance".  The order is listed by the row. In the first, for instance, the order is raw then calibrated.  The ``-- --" indicates those times when both raw and calibrated showed negative trends; ``-- +" indicates when raw data had a negative trend and the calibrated had a positive, and so on.  The capital ``$P$" count large p-values, and the lower case $p$ indicates p-values less than 0.05.

There were $40+36=76$ (26\%) times the two data sets, raw and calibrated, did not agree on the sign of the trend, and $12+37=49$ (17\%) times they did not agree on significance"  This same pattern emerged, with small differences in the counts, for the raw-calibrated-plus-imputed data.  Finally, we again see the imputation process makes a difference, as there were 31 instances of sign differences, and 15 differences in significance.   These findings prove evidence of ``trends" is not as strong as has been previously thought.

{\small
\begin{table}[htbp]
  \begin{tabular}{cccc c cccc}
\multicolumn{9}{c}{Raw - Calibrated}\\
-- -- & -- + & + -- & + +   &   &   $P$ $P$ & $P$ $p$ & $p$ $P$ & $p$ $p$  \\\hline\hline
83 & 40 & 36 & 135 &   & 230 & 12 & 37 & 15\\

\multicolumn{9}{c}{Raw - Imputed}\\
-- -- & -- + & + -- & + + & & $P$ $P$ & $P$ $p$ & $p$ $P$ & $p$ $p$\\\hline\hline
76 & 47 & 24 & 147 && 230 & 12 & 36 & 16 \\

\multicolumn{9}{c}{Calibrated - Imputed}\\
-- -- & -- + & + -- & + + && $P$ $P$ & $P$ $p$ & $p$ $P$ & $p$ $p$\\\hline\hline
94 & 25 & 6 & 169 && 259 & 8 & 7 & 20 \\

  \end{tabular}
\caption{Two-by-two tables of the counts of positive (+) and negative (-) trends, and significant (p) and not-significant (P) trends (i.e. large versus small $p$), for the three comparisons of MAN data. The order of the symbols is Raw then Calibrated (top rows), Raw then Imputed and calibrated (middle rows), and Calibrated and Imputed (bottom rows). It is clear that for some trends identified as negative or significant, and vice versa, in the Raw MAN data were reversed (positive or not-significant) in the Calibrated or Calibrated and Imputed MAN data.}
\label{tab2}
\end{table}
}

The most in-depth (and admittedly tedious) comparison is in Table \ref{tab3}. The designators and orders of trend sign are the same as in Table \ref{tab2}. The difference is that the trends are further broken down by non-significant ($P$) and significant ($p$) trends.  This is done because it is frequently believed trends are not there unless they are significant. Therefore the most important cells in the Table are when significance ``switches" between data sets {\it and} the trend sign changes.  

{\small
\begin{table}[htbp]
  \begin{tabular}{cccc c cccc}
\multicolumn{4}{c}{Raw $P$ - Calibrated $P$} && \multicolumn{4}{c}{Raw $p$ - Calibrated $P$}\\
-- -- & -- + & + -- & + +   & $|$  &   -- -- & -- + & + -- & + +    \\\hline\hline
71 & 40 & 30 & 89 & $|$  & 4 & 0 & 3 & 30\\
\multicolumn{4}{c}{Raw $P$ - Calibrated $p$} && \multicolumn{4}{c}{Raw $p$ - Calibrated $p$}\\
-- -- & -- + & + -- & + +   & $|$  &   -- -- & -- + & + -- & + +    \\\hline\hline
4 & 0 & 3 & 5 & $|$  & 4 & 0 & 0 & 11\\

\\\hline

\multicolumn{4}{c}{Raw $P$ - Imputed $P$} && \multicolumn{4}{c}{Raw $p$ - Imputed $P$}\\
-- -- & -- + & + -- & + +   & $|$  &   -- -- & -- + & + -- & + +    \\\hline\hline
64 & 47 & 22 & 97 & $|$  & 5 & 0 & 2 & 29\\
\multicolumn{4}{c}{Raw $P$ - Imputed $p$} && \multicolumn{4}{c}{Raw $p$ - Imputed $p$}\\
-- -- & -- + & + -- & + +   & $|$  &   -- -- & -- + & + -- & + +    \\\hline\hline
4 & 0 & 0 & 8 & $|$  & 3 & 0 & 0 & 13\\

\\\hline

\multicolumn{4}{c}{Calibrated $P$ - Imputed $P$} && \multicolumn{4}{c}{Calibrated $p$ - Imputed $P$}\\
-- -- & -- + & + -- & + +   & $|$  &   -- -- & -- + & + -- & + +    \\\hline\hline
82 & 25 & 6 & 146 & $|$  & 5 & 0 & 0 & 2\\
\multicolumn{4}{c}{Calibrated $P$ - Imputed $p$} && \multicolumn{4}{c}{Calibrated $p$ - Imputed $p$}\\
-- -- & -- + & + -- & + +   & $|$  &   -- -- & -- + & + -- & + +    \\\hline\hline
1 & 0 & 0 & 7 & $|$  & 6 & 0 & 0 & 14\\

  \end{tabular}
\caption{Finally, a breakdown of the positive (+) and negative (-) trends identified in each of the MAN data sets, conditioned on whether the trends were both not-significant (pp), one significant and one not-significant (pP), vice versa (Pp), and finally both significant (pp). Once again, the data processing method makes a difference in the size, direction, and significance of the estimated trend. Tracking and accounting for the changes introduced in the calibration and imputation is thus crucial.}
\label{tab3}
\end{table}
}

For example, in the upper right most portion of the Table (Raw $p$ - Calibrated $P$) are those times where the raw data showed significant trends (of either sign) and the calibrated did not. Three of those times the raw data had a positive trend, and the calibrated had a negative (but not significant) trend.  This kind of switch is repeated for every differences. 

The lesson from this and the other Tables is that the data manipulation process matters, and can matter a great deal. Decisions about whether a trend is there or not, when based on significance, change, and thus so would judgments of ``significance".  This proves not only that the calibration and imputation process are important, but given the number of switches, the labeling of trends with simple regression is not robust to these manipulations.  This last isn't surprising when we recall the regression lines did not happen---and couldn't have, given the calibration or imputation give different trends.  The interpretation of statistical trends identifying linear causes being suspect is also upheld.

\subsection{Atmospheric NH3}

If the MAN data, in its calibrated or calibrated plus imputed states, is used to infer atmospheric NH3 levels, then the uncertainty in the calibration and imputation {\it must} be accounted for in any analysis, including trend analysis.  That the MAN data itself evinces a trend is only interesting to the extent MAN data is predictive of actual atmospheric NH3, so that any trend analysis must incorporate the uncertainty introduced in calibration. To be clear: we are not interested in trends identified in any of the three MAN data sets, but we want to know whether there are trends in actual atmospheric ammonia, where the MAN data, after calibration, is a prediction of atmospheric ammonia, as above. This section shows how this is done.

Here is the ordinary regression model form for trend, here with $y$ representing atmospheric ammonia, and $x$ the time. 
\begin{equation}
\label{cal1.redux}
y = \theta_0 + \theta_1x + \epsilon,
\end{equation}
where $\theta_1$ is the trend coefficient, with $\epsilon\sim(0,\sigma^2_{\epsilon})$. With the MAN data we do not see $y$, but $y=y'+\nu$ where $y'$ is the MAN data, and $y$ the true value of NH3, which is assumed different by the amount $\nu$, an ``error" or uncertainty introduced by calibration. Thus calibration can be seen as a standard measurement-error regression model; see \cite{CarRup2006}. We assume as is usual $\nu\sim\mbox{N}(0,\sigma^2_{\nu})$, where the normality is an assumption additionally supported by results in Section \ref{scali}. In that Section we learned that the error of $y$ varied with MAN levels, but here we make the conservative simplification of assuming it is constant.   The value of $\sigma^2_{\nu}$ is estimated from the calibration equation of all MAN calibrated data on all LML data, regardless of station. This could, of course, be done on a station-by-station basis, but presumably without much gain in accuracy, given the evidence of the pictures that the individual stations looked similar. In any case, from that regression we have $\hat{\sigma}^2_{\nu} = 1.635$.  The regression equation for trend in MAN data then becomes
\begin{equation}
\label{calerr}
y = \theta_0 + \theta_1x + e,
\end{equation}
(notice the $e$ replaces $\epsilon$) which produces the same estimates of $\theta_0$ and $\theta_1$, but with adjusted standard errors because of the measurement error of $y$.  Let $e\sim\mbox{N}(0,\sigma^2_e)$ as is usual in regression; thus $\sigma^2_e = \sigma^2_{\epsilon} + \sigma^2_{\nu}$. 

\begin{figure}[htbp]
        \centering
        \includegraphics[scale=.4]{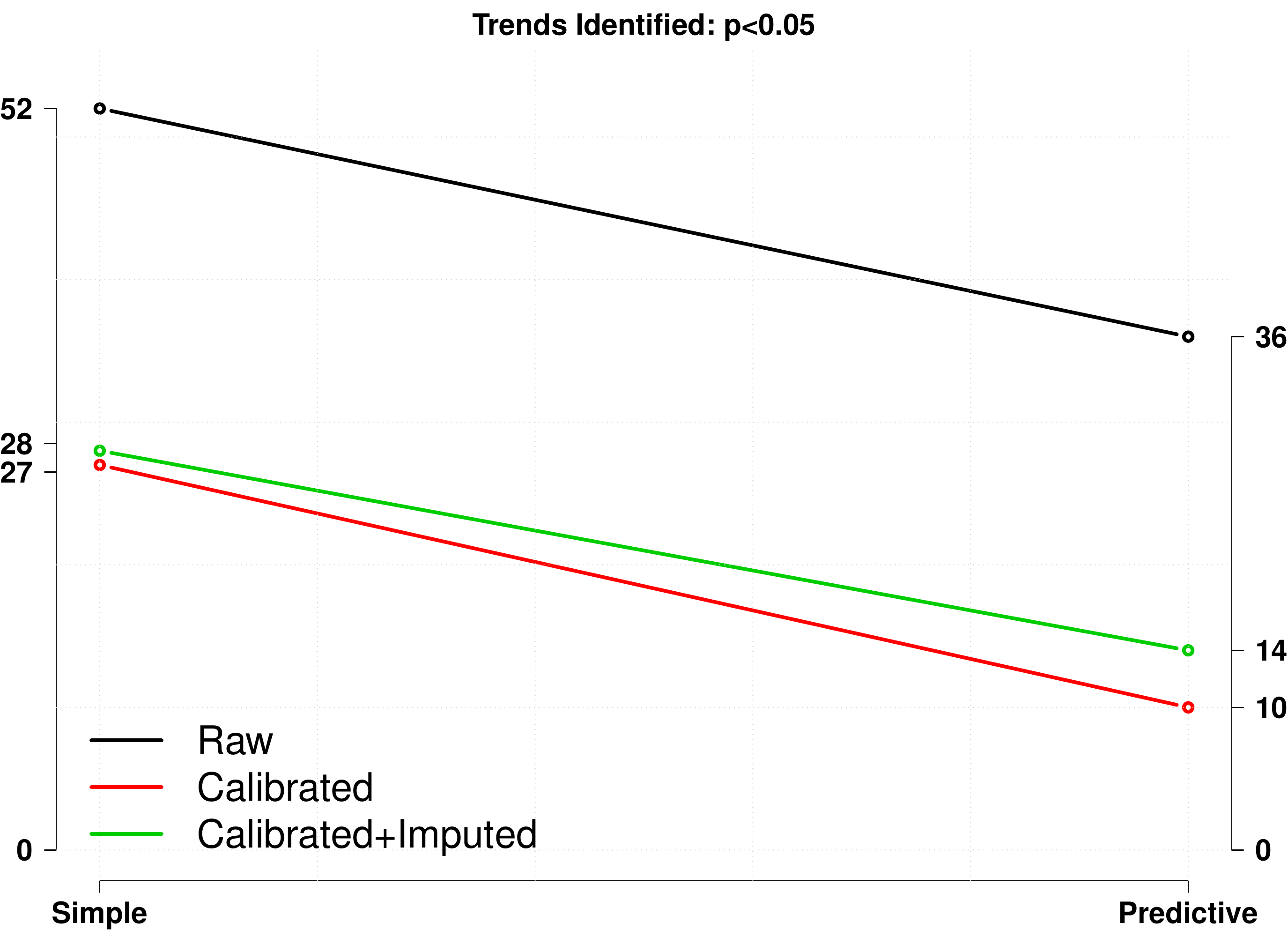}
\caption{\label{figtab4}  The number of significant trends identified in the simple and more complete predictive analysis. The number of trends identified drop by about 50\%.}
\end{figure}

Now p-values in a regression are based on the parameter estimates divided by the standard error of the estimates (this gives the $t$-statistic, from which the p-value is calculated). We recalculated the trend regressions for each MAN data set, computing the new p-values based on $\sigma^2_e$ and compared them to the original p-values based on $\sigma^2_{\epsilon}$. Since $\sigma^2_{\nu}>0$ these new p-values are necessarily always larger. Again, the parameter estimates $\theta_0$ and $\theta_1$ (including the sign) do not change. Only the p-values change to reflect the uncertainty introduced assuming the calibrated MAN data is a prediction of actual atmospheric ammonia.  Table \ref{tab.p} and fig. \ref{figtab4} shows the changes in trends identified by wee p-value from the simple (ignoring the introduced by calibration) to the predictive regressions (incorporating this uncertainty).

{\small
\begin{table}[htbp]
  \begin{tabular}{lcc}
  \multicolumn{3}{c}{Trends identified}\\
Data & Simple & Predictive \\\hline\hline
Raw & 52 & 36\\
Calibrated & 27 & 10 \\
Calibrated+Imputed & 28 & 14 \\
  \end{tabular}
\caption{The number of significant trends identified in the three different data sets, incorporating the uncertainty in the MAN data predicting atmospheric ammonia. This necessarily leads to higher p-values and fewer identified trends.}
\label{tab.p}
\end{table}
}


The significant trends are always fewer, as expected, in the predictive models. Since decisions change based on whether a trend is said to be present or not, the number of decisions that would change are also quite large. The reductions in identified trends are from 33\% to 63\%. Meaning decisions made without using the predictive uncertainty are themselves far too certain. 

It's still worse that it might seem, because the imputation process itself carries uncertainty that should also be incorporated. Imputation is, of course, another way of guessing $y$. We began with calibration with a guess of $y$, which is $y'$. If the imputation process is unbiased in the statistical sense, it would be as if we viewed not $y$ or $y'$, but $y= y'' + \nu+\tau$, where $\tau\sim\mbox{N}(0,\sigma^2_{\tau})$, at least for the points which are imputed. The result is that the p-values would again necessarily increase, and decisions whether trends are present would again change. How much depends on the variance added by the imputation process which we are here unable to estimate, as the imputed data was presented as is.

\section{Conclusion}

In this contribution, we have shown that great over-certainty exists in current methods of trend identification in the LML and MAN datasets. This over-certainty results in part from not tracking uncertainty in the calibration process, which we have done here for the first time. Apart from the calibration process introducing changes and uncertainty in the MAN data, the imputation process further adds to both. We did not reproduce the imputation process itself, but found the process to increase uncertainty in the calibrated data. 

As a result, conclusions about whether trends are present, given the statistical definition of trend used above, are very much dependent on what data source, calibration method, and imputation process are used. Also, the use of p-values of whatever kind does not provide us with insights into the reality of found trends. Worse, the use of p-values as proposed by the RIVM does not conform to the understanding and severe limitations of this statistical entity. In order to really understand trends in datasets such as MAN and LML, uncertainty needs to be accounted for through the whole process of calibration and imputation. This is important, as national decision making is informed by analyses done by the RIVM that lack the critical reflections we have presented and discussed in this contribution.

\bibliographystyle{abbrv}
\bibliography{logic.bib}

\end{document}